\documentclass[conference]{IEEEtran}
\usepackage{amssymb}
\usepackage{mathbbol}
\usepackage{graphicx,cite,epsfig,amssymb,amsmath}
\usepackage{pifont}
\usepackage{bm}
\usepackage{stmaryrd}
\usepackage{subfigure}
\usepackage{bbding}
\usepackage{subfigure}
\usepackage[ruled,vlined]{algorithm2e}
\usepackage[dvipsnames,usenames]{color}
\usepackage{booktabs}

\begin{document}
\title{Progressive Bit-Flipping Decoding of Polar Codes Over Layered Critical Sets}
\author{\IEEEauthorblockN{Zhaoyang~Zhang$^{1,2}$\IEEEauthorrefmark{2}, Kangjian~Qin$^{1,2}$, Liang~Zhang$^{1,2}$, Huazi~Zhang$^3$, Guo~Tai~Chen$^4$}
\IEEEauthorblockA{$^1$College of Information Science and Electronic Engineering, Zhejiang University, Hangzhou, China\\
$^2$Zhejiang Provincial Key Laboratory of Info. Proc., Commun. \& Netw. (IPCAN), Hangzhou, China\\
$^3$Huawei Technologies Co., Ltd., Hangzhou, China\\
$^4$School of Electrical and Information Engineering, Fuqing Branch of Fujian Normal University, Fuzhou, China\\
E-mail: \{ning\_ming\IEEEauthorrefmark{2}, north\}@zju.edu.cn, 0705zhangliang@sina.com, tom.zju@gmail.com, chenguot@163.com}}

\maketitle
\begin{abstract}
In successive cancellation (SC) polar decoding, an incorrect estimate of any prior unfrozen bit may bring about severe error propagation in the following decoding, thus it is desirable to find out and correct an error as early as possible. In this paper, we first construct a critical set $S$ of unfrozen bits, which with high probability (typically $>99\%$) includes the bit where the first error happens. Then we develop a progressive multi-level bit-flipping decoding algorithm to correct multiple errors over the multiple-layer critical sets each of which is constructed using the remaining undecoded subtree associated with the previous layer. The \emph{level} in fact indicates the number of \emph{independent} errors that could be corrected. We show that as the level increases, the block error rate (BLER) performance of the proposed progressive bit flipping decoder competes with the corresponding cyclic redundancy check (CRC) aided successive cancellation list (CA-SCL) decoder, e.g., a level 4 progressive bit-flipping decoder is comparable to the CA-SCL decoder with a list size of $L=32$. Furthermore, the average complexity of the proposed algorithm is much lower than that of a SCL decoder (and is similar to that of SC decoding) at medium to high signal to noise ratio (SNR).
\end{abstract}

\section{introduction}

Polar codes, as the first provable capacity-achieving codes for any symmetric binary-input discrete memoryless channel (B-DMC) with efficient successive cancellation (SC) decoding \cite{Channel_polarization_Arikan}, have been recently adopted as the channel coding scheme for control information in the 5G enhanced Mobile BroadBand (eMBB) scenario \cite{3GPP_5G_NR}. Different from data packets, the block-length for control messages is typically short or moderate due to coding granularity. However, the performance of such finite block-length polar codes is still far from satisfactory.

To improve the performance of polar codes in finite block-length case, Tal and Vardy presented a successive cancellation list (SCL) decoder in \cite{list_decoding_Vardy}, which helps polar codes successfully compete with low-density parity-check (LDPC) codes. Subsequently, adaptive SCL decoding and cyclic redundancy check (CRC) aided SCL (CA-SCL) decoding were proposed in \cite{An_adaptive_Li,Improved_successive_Chen}. Moreover, the performance of SCL decoding was theoretically analyzed in \cite{Scaling_exponent_Mondelli}. Although SCL decoder significantly improves the block error rate (BLER) of finite block-length polar codes, it suffers from large storage overhead and high computational complexity, both of which grow linearly with the list size. To address this issue, the authors in \cite{Low-complexity_improved_SCL_Orion_Afisiadis} put forward a SC flip decoder trying to correct the first erroneous estimate of an unfrozen bit, and indicated that the decoding performance could be dramatically improved if the first incorrect hard decision was flipped. This decoder was further modified in \cite{An_Improved_SCFlip_Decoder_for_PC} to recover two incorrect hard decisions, which induced significant gains in terms of decoding performance and competed with the CA-SCL decoder with list size $L=4$. Furthermore, \cite{An_Improved_SCFlip_Decoder_for_PC} defined a new metric to determine the flipping positions, which yielded reduced complexity compared to the log likelihood ratio (LLR) metric exploited in \cite{Low-complexity_improved_SCL_Orion_Afisiadis}. Nonetheless, by using such metric, the search scope for the first erroneous hard decision is still the entire unfrozen set.

In this paper, by investigating the distribution of the first erroneous hard decision in SC decoding, we find it possible to narrow down the search scope to an unfrozen bit subset $S$, which is much smaller than the unfrozen set. For ease of exposition, the subset $S$ is referred to as \emph{critical set} through the rest of this paper. It can be proven that if SC decoding fails, the first incorrect hard decision is almost surely included this critical set. As such, the decoder only needs to consider $S$ for the flipping position, thus further reducing the computational complexity. In addition, since there might exist several other errors besides the first erroneous hard decision, it is desirable to flip multiple incorrect bits rather than only the first one. For this purpose, we propose to iteratively modify the critical set $S$ and correct the errors progressively, aiming to achieve superior decoding performance. Numerical results show that, the proposed progressive decoder can compete with the CA-SCL decoder in terms of BLER performance, e.g., a level 4 progressive bit-flipping decoder is comparable to the CA-SCL decoder with a list size of $L=32$, while having an average decoding complexity similar to that of the standard SC decoding at medium to high SNR.

To summarize, our main contributions are as follows:
\begin{itemize}
    \item The critical set, which with high probability includes the first incorrect hard decision, is proposed. Because of the smaller search scope, the computational complexity is reduced significantly.
    \item A progressive multi-level bit-flipping decoding algorithm based on iteratively modified critical set is proposed. It has the ability to correct multiple errors and achieve a BLER performance much better than the conventional SC decoding and comparable to the CA-SCL decoding.
\end{itemize}

The rest of this paper is organized as below. In Section \uppercase\expandafter{\romannumeral2}, a short background on polar codes is presented and our analytical framework is briefly described. Section \uppercase\expandafter{\romannumeral3} provides some important results about the critical set. Section \uppercase\expandafter{\romannumeral4} shows the proposed progressive algorithm that correct multiple errors. Simulation results are provided in Section \uppercase\expandafter{\romannumeral5} and Section \uppercase\expandafter{\romannumeral6} concludes this paper.

\section{Preliminaries}
\subsection{Polar codes}
We use $a_1^N$ to denote a vector $(a_1,a_2,...,a_N)$. For polar codes with block-length $N=2^n$ and kernel $\tiny{\bm{G}_2=\begin{bmatrix} 1\;0 \\ 1\;1 \end{bmatrix}}$, we denote $u_1^N$ as the information sequence, and a polar codeword $c_1^N$ is obtained by $c_1^N=u_1^N\bm{B}_N\bm{G}_2^{\otimes n}$, where `$\otimes$' denotes the Kronecker product while $\bm{B}_N$ is a permutation matrix. A coding rate $R=K/N$ means that a set $\mathcal{A} \subset \{1,2,...,N\}$ of cardinality $K$ is selected as the information set (see \cite{Channel_polarization_Arikan}), and thus $u_1^N$ consists of $K$ unfrozen bits and $N-K$ frozen bits (all frozen bits are assumed to be zero if not specified). The split channel is defined as $W_N^{(i)}(y_1^N,u_1^{i-1}|u_i)=\sum_{u_{i+1}^N \in \mathcal{X}^{N-i}}\frac{1}{2^{N-1}}W_N(y_1^N|u_1^N)$, and the Bhattacharyya parameter $Z(W_N^{(i)})$ is computed to select the $K$ most reliable split channels to transmit unfrozen bits. Interested readers are referred to \cite{Channel_polarization_Arikan} for more details.

\subsection{Analytical framework}
The framework in \cite{A_simplified_Alamdar} is adopted for the ensuing analysis. To facilitate understanding, let us consider a toy example of polar codes with block-length $N=2^{2}$ and information sequence $u_{1}^{4}=\left(u_{1},u_{2},u_{3},u_{4}\right)$. $u_{3}$ and $u_{4}$ are chosen as the unfrozen bits, thus inducing a coding rate $R=2/4$.

\begin{figure}[!ht]
\centering
\includegraphics[scale=0.5]{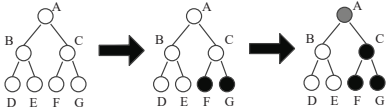}
\caption{Full binary tree for $N=2^2$.}
\label{binary_tree_toy_example}
\end{figure}

To proceed, a full binary tree with $N=2^{2}$ leaf nodes is constructed in Fig. \ref{binary_tree_toy_example} (left). The leaf nodes $\{D,E,F,G\}$ correspond to the information bits $\{u_1,u_2,u_3,u_4\}$, respectively. Since $u_{3}$ and $u_{4}$ are unfrozen bits, nodes $F$ and $G$ are denoted by black circles for the sake of clarity, see Fig. \ref{binary_tree_toy_example} (middle) for illustration. Furthermore, for each non-leaf node in the tree, if its two descendants are of the same color, then it is marked with that color as well. Otherwise, it is indicated by a gray circle. This process starts from the bottom non-leaf nodes until the root node is reached, as shown in Fig. \ref{binary_tree_toy_example} (right).

To implement polar encoding, a constituent code is assigned to each node in Fig. \ref{binary_tree_toy_example} (right). Suppose that $u_1^4=(0,0,1,0)$, i.e., $D[1]=0$, $E[1]=0$, $F[1]=1$, $G[1]=0$, where $D[i]$ denotes the $i$-th component at node $D$. On this basis, the constituent code at node $B$ is obtained by $(B[1],B[2])=(D[1],E[1])\times \bm{G}_2$, which gives $B[1]=B[2]=0$. Similarly, we have $C[1]=1$ and $C[2]=0$. Next, invoking the expressions $(A[1],A[2])=(B[1],C[1])\times \bm{G}_2$ and $(A[3],A[4])=(B[2],C[2])\times \bm{G}_2$, the polar codeword is obtained as $c_1^4=(A[1],A[2],A[3],A[4])=(1,1,0,0)$. One can also check that $c_1^4$ can be obtained by $c_1^4=(D[1],E[1],F[1],G[1])\times \bm{B}_4\bm{G}_2^{\otimes 2}$. As for SC decoding, it starts from the root node $A$, which possesses the LLRs received from the underlying channel, and uses\cite[Eq. 75]{Channel_polarization_Arikan} and \cite[Eq. 76]{Channel_polarization_Arikan} to calculate LLRs recursively. In the meantime, polarization can also be interpreted based on this tree. One can check that node $A$ has four independent copies of the underlying channel $W$, while node $B$ has two independent copies of the synthetic channel $W_2^{(1)}$ and node $C$ has two independent copies of the synthetic channel $W_2^{(2)}$. Finally, a leaf node has a unique copy of the split channel, e.g., node $E$ has $W_4^{(2)}$. We refer the reader to \cite{A_simplified_Alamdar} for more details. It is worth noting that this framework can be extended to polar codes with arbitrary block-length.

\section{The critical set}
\subsection{SC decoding from a subblock-by-subblock perspective}

\begin{figure}[!ht]
\centering
\vspace{-0.2cm}
\includegraphics[scale=0.5]{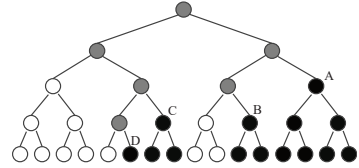}
\caption{Full binary tree for $N=2^4$.}
\label{binary_tree_complicated_example}
\end{figure}

Let us focus on a more complicated example as shown in Fig. \ref{binary_tree_complicated_example}, where $N=2^4$ and $K=9$ (the information set may not be reasonable).

In our framework, SC decoding is not viewed as a bit-by-bit process, but from a subblock-by-subblock perspective. Once the full binary tree corresponding to the current specific polar code is constructed, the entire polar code is divided into multiple sub-polar codes (also called subblocks), which all have coding rate $R=1$. In Fig. \ref{binary_tree_complicated_example}, there exist four such subblocks, which are denoted by the corresponding root nodes $A$, $B$, $C$ and $D$, respectively (they are also polar codes but with shorter block-length). The subblock consists of only unfrozen bits, e.g, node $A$ has unfrozen bits $u_{13}^{16}$. In particular, node $D$ has an unfrozen bit $u_6$, and it can be viewed as a special subblock which has itself as both the codeword (root node) and information sequence (leaf node).

Now, consider a general subblock $A$ ($A$ denotes its root node) which has $M=2^m$ unfrozen bits. We use $u_1^M$ and $c_1^M$ to denote its information sequence and codeword, respectively. Then the following proposition is derived, which sheds light on our main results.

\newtheorem{proposition}{Proposition}
\begin{proposition}
For a binary erasure channel (BEC), the entire subblock is correctly decoded if and only if $u_1$ is correctly decoded.
\end{proposition}
\begin{IEEEproof}
The proof is straightforward. Recall that $u_1^M=c_1^M(\bm{B}_M\bm{G}_2^{\otimes m})^{-1}$, and one can check that we always have $u_1=c_1\oplus c_2\oplus \cdots \oplus c_M$. If $u_1$ is correctly decoded, then it means that there must be no erasure symbols involved in $c_1^M$, and thus $u_1^M=c_1^M(\bm{B}_M\bm{G}_2^{\otimes m})^{-1}$. On the other hand, if the entire subblock is correctly decoded, i.e., every $c_i$ takes a value either $0$ or $1$, it is obvious that $u_1$ can be correctly estimated as well.
\end{IEEEproof}

Now, we extend the above arguments to other channels. According to our framework, node $A$ has $M=2^m$ independent copies of some synthetic channel, which is denoted by $W_M$, and we further denote the split channel experienced by $u_i$ (within this subblock) as $W_M^{(i)}$. Provided that all the prior subblocks are correct, we assume the error probabilities of $W_M$ and $W_M^{(i)}$ are $p$ and $P_{u_i}$, respectively. Under this condition, we obtain the following proposition.

\begin{proposition}
Denote the error probability of the entire subblock as $P_{\text{s\_bler}}$. Then, for $p<\epsilon$, we have $P_{\text{s\_bler}}-P_{u_1}<\sum_{i=1}^{M/2}C_{M}^{2i}{\epsilon}^{2i}$.
\end{proposition}

\begin{IEEEproof}
According to the number of errors occurred in the codeword, $P_{\text{s\_bler}}$ can be computed as $P_{\text{s\_bler}}=C_{M}^{1}p(1-p)^{M-1}+C_{M}^{2}p^2(1-p)^{M-2}+\cdots +C_{M}^{M}p^M$. Although this is not BEC, however, no frozen bits are involved, and thus no parity check needs to be satisfied. Then, the estimate of $u_1$, denoted as $\hat{u}_1$, can still be computed by $\hat{u}_1=\hat{c}_1\oplus \hat{c}_2\oplus \cdots \oplus \hat{c}_M$, where $\hat{c}_i$ denotes the hard decision. Thus, $u_1$ is incorrectly decoded if and only if the number of errors in $c_1^M$ is odd, which gives $P_{u_1}=C_{M}^{1}p(1-p)^{M-1}+C_{M}^{3}p^3(1-p)^{M-3}+\cdots +C_{M}^{M-1}p^{M-1}(1-p)$. As such, it is obtained that
\begin{equation}\nonumber
\begin{aligned}
P_{\text{s\_bler}}-P_{u_1}=\sum_{i=1}^{M/2}C_{M}^{2i}p^{2i}(1-p)^{M-2i}<\sum_{i=1}^{M/2}C_{M}^{2i}{\epsilon}^{2i},
\end{aligned}
\end{equation}
which completes the proof.
\end{IEEEproof}

\emph{Remarks:} The difference $P_{\text{s\_bler}}-P_{u_1}$ represents the probability that two or more errors occur. As $\epsilon \rightarrow 0$, this value approaches $0$. This implies that if $W_M$ is reliable enough, $P_{u_1}$ is quite close to the error probability of the entire subblock.

\subsection{Constructing the critical set}

Capitalizing on the results above, there is a high probability that the first incorrectly estimated unfrozen bit happens to be the first unfrozen bit within the subblock. Inspired by this, we provide a method to construct a set $S$ that almost surely includes the first incorrect hard decision in SC decoding. The corresponding algorithm is summarized as Algorithm \ref{alg:A}.
\begin{algorithm}
\caption{A method to construct the critical set $S$}
\label{alg:A}
\begin{enumerate}
\item[Step 1] Establish the full binary tree corresponding to the current polar codes;
\item[Step 2] Divide the polar codes into multiple subblocks with coding rate $R=1$ and put the first unfrozen bit of each subblock into set $S$.
\end{enumerate}
\end{algorithm}

Taking Fig. \ref{binary_tree_complicated_example} for instance, we have $S=\{u_6,u_7,u_{11},u_{13}\}$. Note that the number of elements in set $S$ is exactly the same as that of subblocks, which is rather small compared with the information set $\mathcal{A}$ of cardinality $K$. Furthermore, $S$ is uniquely determined once the construction of polar codes is completed.
\subsection{Validation of set $S$ under Gaussian approximation}
To validate the above method, we first focus on the evaluation of the difference $P_{\text{s\_bler}}-P_{u_1}$. To the best of our knowledge, the exact values of $P_{\text{s\_bler}}$ and $P_{u_1}$ are rather difficult to compute. Thereby, we exploit the Gaussian approximation method to provide some insightful results. Gaussian approximation was introduced in \cite{Analysis_of_sum-product_Richardson} and adopted for the analysis of polar codes in \cite{Efficient_design_Trifonov}. In the following analysis, we restrict our attention to binary phase shift keying (BPSK) modulation, i.e., for a given AWGN channel, the received symbol is expressed as $y_i=x_i+n_i$, where $x_i=1-2c_i$, and $n_i$ represents a Gaussian random variable with mean zero and variance $\sigma^2$. Without loss of generality, we assume that the all-zero codeword is transmitted. In this sense, one can check that the received LLR can be written as $L(y_i)=\mathrm{log}\frac{W(y_i|x_i=1)}{W(y_i|x_i=-1)}=\frac{2y_i}{\sigma^2}$, which can be viewed as a Gaussian random variable with mean $\frac{2}{\sigma^2}$ and variance $\frac{4}{\sigma^2}$. By rewriting formulae \cite[Eq.75]{Channel_polarization_Arikan} and \cite[Eq.76]{Channel_polarization_Arikan} in an LLR form, one can find that the operations involved in SC decoding are exactly the same as those in belief propagation decoding. Thus, as suggested in \cite{Analysis_of_sum-product_Richardson}, by assuming that the symmetry condition is always satisfied, all the LLRs involved in SC decoding can be viewed as Gaussian random variables with the form $\mathcal{N}(\mu,2\mu)$. We only need to calculate the mean $\mu$.

For a given subblock $A$, suppose that the LLR corresponding to the synthetic channel $W_M$ satisfies $\mathcal{N}(\mu,2\mu)$. Then the LLR corresponding to the split channel $W_M^{(1)}$ takes a mean $\mu_{u_1}=\phi^{-1}(1-\big(1-\phi(\mu)\big)^{2^m})$, where $\phi(x)$ is defined as
\begin{eqnarray}\nonumber
& \phi(x)=\left\{
 \begin{aligned}
    &1-\frac{1}{\sqrt{4 \pi x}}\int_{-\infty}^{\infty}\mathrm{tanh}\frac{u}{2}\cdot e^{-\frac{(u-x)^2}{4x}}du, \quad x> 0,\\
    &1, \quad  \quad \quad\quad\quad\quad\quad\quad\quad\quad\quad\quad\quad\quad\quad~~ x=0.\\
 \end{aligned}
 \right.
\end{eqnarray}
Due to the all-zero codeword, the probability that $u_1$ is incorrectly estimated is calculated as $P_{u_1}=Q(\sqrt{\mu_{u_1}/2})$,
where $Q(x)=\frac{1}{\sqrt{2\pi}}\int_x^{+\infty}e^{-\frac{t^2}{2}}dt$. Similarly, the error probability of the entire subblock is $P_{\text{s\_bler}}=1-(1-Q(\sqrt{\mu/2}))^{2^m}$.

\begin{figure}[!ht]
\centering
\includegraphics[scale=0.46]{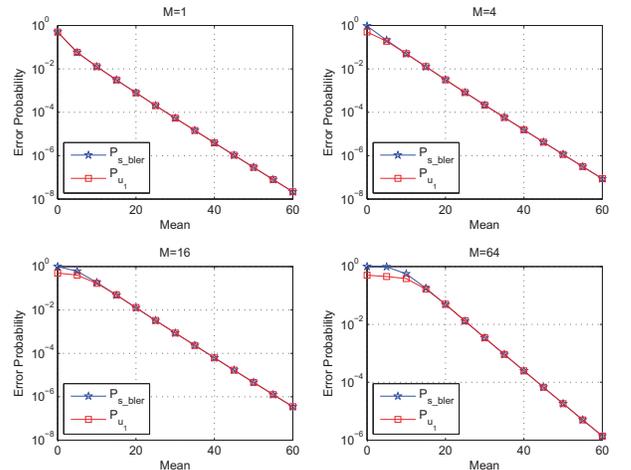}
\caption{$P_{\text{s\_bler}}$ vs. $P_{u_1}$ under Gaussian approximation.}
\label{GA_results}
\end{figure}

The numerical comparison between $P_{\text{s\_bler}}$ and $P_{u_1}$ using Gaussian approximation is depicted in Fig. \ref{GA_results}. It can be observed that for large length $M$ and small $\mu$, there still exists some obvious difference between $P_{\text{s\_bler}}$ and $P_{u_1}$. This is practical because, firstly, small $\mu$ means that the synthetic channel $W_M$ is not quite reliable, and thus it is more likely to introduce more than one error; secondly, large $M$ increases the probability to include two or more errors as well. Thus, the difference $P_{\text{s\_bler}}-P_{u_1}=\sum_{i=1}^{M/2}C_{M}^{2i}p^{2i}(1-p)^{M-2i}$ becomes noticeable. However, we conjecture that for any given underlying channel $W$ to implement polarization, a subblock $A$ with large $M$ in general has a large $\mu$ as well. For larger $M$, the split channel $W_M^{(1)}$ is further degraded compared with the synthetic channel $W_M$. As $W_M^{(1)}$ is selected to transmit an unfrozen bit, thus $W_M$ should be sufficiently reliable as well, since otherwise $u_1$ will turn into a frozen bit. It can be seen in Fig. \ref{GA_results} that if $\mu$ is large enough, the difference between $P_{\text{s\_bler}}$ and $P_{u_1}$ can usually be neglected.

\begin{table}[ht]
\begin{center}
\caption{EVALUATION of ALGORITHM 1}
\label{table1}
\begin{tabular}[h]{|c|c|c|c|c|c|}
\hline
\multicolumn{6}{|c|}{$N=1024,K=512$, simulation times $T=10^6$}\\
\hline
$E_b/N_0$(dB) & 1 & 1.5 & 2 & 2.5 &3 \\
\hline
Included in set $S$ & 675840 & 296391& 73789 & 10888 &1007 \\
\hline
Incorrect blocks & 677211 & 296573 & 73810 & 10888 & 1007 \\
\hline
Accuracy ($\%$) & 99.80 & 99.94 & 99.97 & 100 & 100 \\
\hline
Size of set $S$ & 110 & 112 & 117 & 124 & 129 \\
\hline
\end{tabular}
\end{center}
\vspace{-0.4cm}
\end{table}

To further evaluate the Algorithm \ref{alg:A}, we focus on the probability that the first incorrect hard decision falls into set $S$ through Monte Carlo simulations, which is shown in Table \ref{table1}. The ``Included in set $S$'' denotes the number that the first incorrectly estimated unfrozen bit falls into the critical set $S$, and ``Incorrect blocks'' denotes the total number of blocks that are not correctly recovered, while ``Accuracy'' simply computes their ratio. It can be observed that the probability that the first error is included in the critical set $S$ approaches $100\%$, even for low signal to noise ratios (SNRs). Furthermore, the performance of Algorithm 1 improves as SNR increases, which is consistent with our prior analysis.

\section{Progressively correcting multiple errors}
In this section, invoking the derived critical set, the bit-flipping methodology is adopted to correct the first erroneous unfrozen bit. Furthermore, by iteratively modifying the critical set, a progressive multi-level bit-flipping decoding algorithm, which can correct multiple errors in SC decoding, is proposed.

\subsection{Progressive bit-flipping decoding }
Now suppose that $\hat{u}_i$ is the first incorrect hard decision and is flipped to $1-\hat{u}_i$ based on the bit-flipping method. Under this condition, all the elements in $u_1^i$ can be viewed as frozen bits. The reason is that, for the split channel $W_N^{(i+1)}(y_1^N,u_1^i|u_{i+1})$, the estimate of $u_{i+1}$ is determined once the sequence $\hat{u}_1^i$ is provided. And therefore, whether some $u_j$ with $1\le j \le i$ is a frozen bit or unfrozen bit no longer makes any difference.

By considering $\{u_1,\cdots,u_i\}$ as frozen bits, a new full binary tree similar to Fig. \ref{binary_tree_complicated_example} can be established immediately. However, at this time, all nodes corresponding to $u_1^i$ are white nodes, while the colors of the following nodes corresponding to $u_{i+1}^N$ still depend on whether it is a frozen bit or unfrozen bit. Based on this new tree, we can construct a modified critical set $S'$ using Algorithm \ref{alg:A}. This modified critical set implies that if errors occur in estimating $u_{i+1}^N$ under SC decoding, the first incorrect hard decision should be almost surely included in such a set. By adopting the bit-flipping operation as done for $u_i$, this error is promised to be corrected, thus further improving the performance. Note that, including $u_i$, the above scheme has corrected two errors during SC decoding.

\begin{figure}[!ht]
\centering
\includegraphics[scale=0.6]{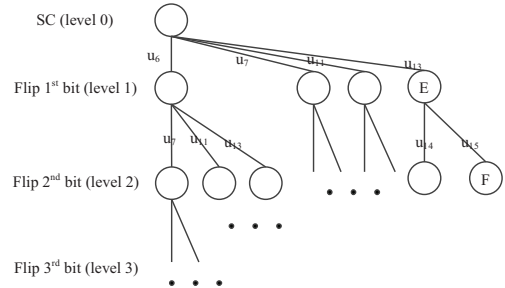}
\caption{Implementation of progressive bit-flipping decoding for Fig. \ref{binary_tree_complicated_example}.}
\label{treestructure}
\end{figure}

Interestingly, this scheme can be extended to correct more errors based on a tree structure. Taking the polar code in Fig. \ref{binary_tree_complicated_example} for example, the tree structure based implementation of progressive bit-flipping decoding is depicted in Fig. \ref{treestructure}, where each node denotes an estimate of $u_1^N$ as a candidate sequence and the edge indicates the unfrozen bit that is flipped. The tree is built via the following steps: first, conventional SC decoding is employed to obtain the root node at level $0$, which denotes the candidate sequence $\hat{u}_1^N$ without flipping any bits; next, the critical set $S_1=\{u_6,u_7,u_{11},u_{13}\}$ is constructed to obtain nodes at level $1$, i.e., every unfrozen bit in $S_1$ corresponds to an edge extended from the root node; then, for each node at level $1$, e.g., node $E$, it constructs a modified critical set $S'=\{u_{14},u_{15}\}$ by building a full binary tree similar to Fig. \ref{binary_tree_complicated_example}, with the leaf nodes corresponding to $u_1^{13}$ being white and those corresponding to $u_{14}^{16}$ being black, thus inducing the edges and nodes at level $2$. Intuitively, repeating the steps above gives rise to the following levels of the tree.

On the basis of such tree structure, the bit-flipping decoding scheme is implemented in a level-order traversal, starting from the root node. In particular, for each node, the entire edges that start from the root node constitute the unfrozen bits that should be flipped. For instance, the node $F$ in Fig. \ref{treestructure} means that: SC decoding is first implemented to compute $\hat{u}_1^{13}$, but $\hat{u}_{13}$ is flipped; then SC decoding is continued to compute $\hat{u}_{14}^{15}$, but $\hat{u}_{15}$ is flipped as well; finally SC decoding is implemented to compute $\hat{u}_{16}$ and thus a candidate sequence $\hat{u}_1^{16}$ is obtained at this node. The ``level'' in fact indicates the number of unfrozen bits that are flipped, and specifically level $0$ denotes the conventional SC decoding without flipping any unfrozen bits. Therefore, the progressive bit-flipping decoding can be viewed as a tree search process.

\subsection{Pruning technique}

To further reduce the search complexity, the current node should not generate any child node if it contains some incorrectly flipped unfrozen bits. According to Gaussian approximation, if all the prior flipped unfrozen bits are correct, the LLRs at $u_i^{N}$ should not be too small compared with their mean values. Based on this observation, we can assign a threshold $\omega_l$ to the $l$-th level in the tree structure, and design a metric $\mu_i-\gamma_{\text{left}}\sigma_i$, where $\sigma_i=\sqrt{2\mu_i}$ and $\gamma_{\text{left}}$ is an optimized value derived through numerical simulations. By counting the number of unfrozen bits in $u_i^N$ as $N_1$, while the number of unfrozen bits whose LLRs fail in achieving $\mu_i-\gamma_{\text{left}}\sigma_i$ as $N_2$ (note that unfrozen bits belonging to the critical set are excluded when counting $N_1$ and $N_2$), we define $E_\textrm{NoChild}$ as the event $\frac{N_2}{N_1}>=\omega_l$, if $E_\textrm{NoChild}$ is true, then the estimate sequence $\hat{u}_1^{i-1}$ is supposed to contain at least one error, thus the branches extended from the current node can be pruned. Otherwise, the current node is allowed to generate child nodes.

If the current node is determined to generate its child nodes, then the unfrozen bits which are likely to be correct should not be selected as child nodes. Recall that under Gaussian approximation, given that $\hat{u}_1^{i-1}=u_1^{i-1}$, if $\textrm{LLR}(u_i)$ is larger than its mean value, then $u_i$ is supposed to be correct. On this basis, we design a threshold $\mu_i+\gamma_{\text{right}}\sigma_i$, where $\sigma_i=\sqrt{2\mu_i}$ and $\gamma_{\text{right}}$ is a constant. We define $E_\textrm{NotSelect}(u_i)$ as the event $L(u_i)>\mu_i+\gamma_{\text{right}}\sigma_i$, if event $E_\textrm{NotSelect}(u_i)$ is true, then $u_i$ is not selected to be the child node.

The proposed progressive multi-level bit-flipping decoding algorithm using the above pruning rules is summarized as Algorithm \ref{alg:B}, where $S_l$ denotes the set of unfrozen bits that should be flipped at level $l$.
\vspace{-0.2cm}
\begin{algorithm}
\caption{Progressive bit-flipping decoding}
\label{alg:B}
\LinesNumbered 
\KwIn{the received vector $y_1^N$, unfrozen set $\mathcal{A}$}
\KwOut{recovered sequence $\hat{u}_1^{N}$ }
$\hat{u}_1^{N}\leftarrow \textrm{SC}(y_1^N,\mathcal{A})$, $l \leftarrow 0$ \qquad //initialization\\
\While{$\textrm{CRC}(\hat{u}_1^{N})=\textrm{failure}$ }{
　　$l\leftarrow l+1$\\
    generate critical sets at level $l$ to form $S_l$\\
    select some $u_i \in S_l$ in an increasing order of $\frac{|L(u_i)|}{\mu_i}$ \\
    $\hat{u}_1^{N}\leftarrow \textrm{Bit-Flipping}(y_1^N,\mathcal{A},u_i)$\\
    \While{ $\textrm{CRC}(\hat{u}_1^{N})=\textrm{failure}$ }{
    \eIf{$E_\textrm{NoChild}=\textrm{false}$ }{
　　　　construct $S$ of current node\\
　　　　\If{$E_\textrm{NotSelect}(u_i)=\textrm{true}$}{
　　　　remove $u_i$ from $S$\\
        generate child nodes using $S$
　　}}{
   \eIf{every $u_i \in S_l$ has been flipped}{
　　　　go to step 3
　　}{
　　　　go to step 5
　　}}}}
\textbf{return} $\hat{u}_1^{N}$
\vspace{-0.05cm}
\end{algorithm}

\section{Simulation results}
In this section, the BLER performance and the computational complexity of the proposed progressive bit-flipping algorithm are investigated. Specifically, we focus on transmissions with BPSK modulation over AWGN channel (details have been given in Section \uppercase\expandafter{\romannumeral3}-C). Polar codes are constructed with parameters $N=1024$ and $K=512$ using Gaussian approximation as in \cite{Efficient_design_Trifonov} and then concatenated with a $24$-CRC with generator polynomial $g(D)=D^{24}+D^{23}+D^6+D^5+D+1$ (see \cite{Digital_Communications}). In this regard, the coding rate for polar codes is $R=1/2$ while the effective information rate is $R=\frac{K-24}{N}$.

\begin{figure}[!ht]
\centering
\vspace{-0.25cm}
\includegraphics[scale=1]{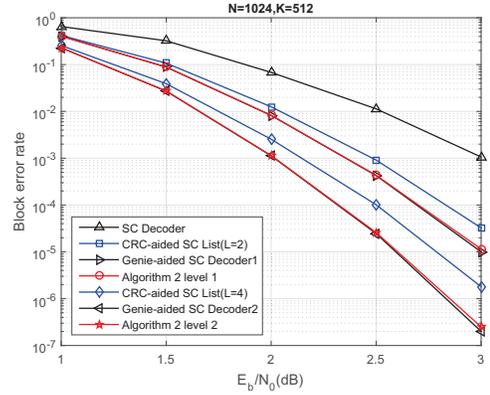}
\caption{BLER performance of Algorithm \ref{alg:B} with level $=1$ and level $=2$.}
\label{performance}
\vspace{-0.25cm}
\end{figure}
\begin{figure}[!ht]
\centering
\vspace{-0.25cm}
\includegraphics[scale=0.5]{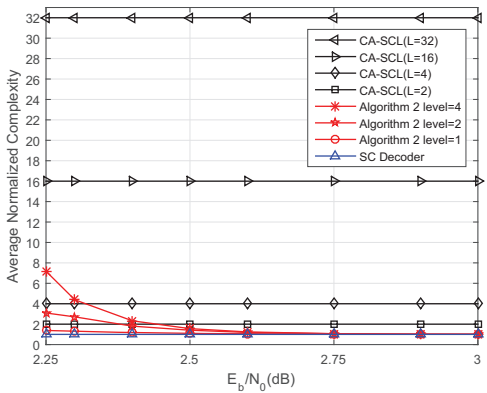}
\caption{Average complexity of Algorithm \ref{alg:B} normalized by the complexity of standard SC decoding}
\label{complexityanalysis}
\vspace{-0.26cm}
\end{figure}

In Fig. \ref{performance}, we compare the BLER performance of Algorithm \ref{alg:B} with $\textrm{level}=\{1,2\}$ and CA-SCL decoder with list size $L=\{2,4\}$.
In particular, the pruning rules introduced in Section \uppercase\expandafter{\romannumeral4}-B are not used here, i.e., each node at $\textrm{level}=\{1,2\}$ always chooses to generate its child nodes. We also use the genie-aided SC decoder (also called Oracle-Assisted SC Decoder), as in \cite{Low-complexity_improved_SCL_Orion_Afisiadis,An_Improved_SCFlip_Decoder_for_PC}, to predict the theoretical optimal performance, which serve as lower bounds on the BLER results for practical SC flip decoders. ``Genie-aided SC Decoder $k$'' means that it can always correct the first $k$ incorrect hard decisions met by SC decoder, but no more errors can be corrected.
As shown in Fig. \ref{performance}, the BLER performance of Algorithm \ref{alg:B} with $\textrm{level}=1$ outperforms the CA-SCL decoder with list size $L=2$, but with only $50\%$ computational complexity at medium to high SNR region (see Fig. \ref{complexityanalysis}), and as the level increases to levle$=2$, Algorithm \ref{alg:B} outperforms the CA-SCL decoder with $L=4$, but with only $25\%$ computational complexity at medium to high SNR region. Furthermore, Algorithm \ref{alg:B} can achieve almost the same performance as the Genie-aided SC decoder if they are designed to correct the same number of incorrect hard decisions in SC decoding.

\begin{figure}[!ht]
\centering
\vspace{-0.25cm}
\includegraphics[scale=0.5]{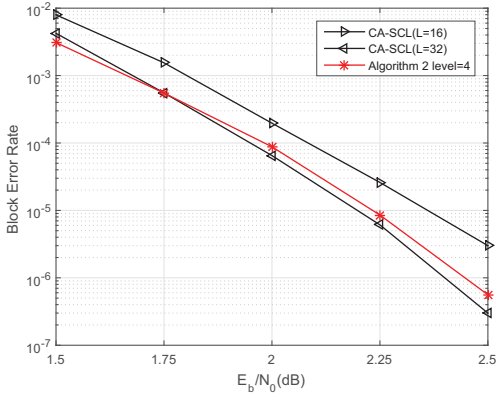}
\caption{BLER performance of Algorithm \ref{alg:B} with level $=4$ vs. CA-SCL decoder with $L=16$ and $L=32$.}
\label{performance2}
\vspace{-0.25cm}
\end{figure}

In Fig. \ref{performance2}, we compare the BLER performance of Algorithm \ref{alg:B} with $\textrm{level}=4$ and CA-SCL decoder with list size $L=\{16,32\}$. The detailed pruning rules and corresponding parameters are listed in Table \ref{table2}, where the $\varnothing$ notation means that the corresponding pruning rule introduced in Section \uppercase\expandafter{\romannumeral4}-B is not used. For instance, $\omega_0$ is $\varnothing$ for all SNRs, which implies that for each node (in fact only one) at level 0, it always chooses to generate its child nodes. We observe that for higher decoding level, such as level $=4$, the proposed bit-flipping decoder can achieve superior BLER performance, which competes with the CA-SCL decoder with a list size $L=32$ and outperforms CA-SCL decoder with $L=16$. Moreover, the computational complexity is dramatically reduced and even degrades to that of SC decoding at medium to high SNR region (see Fig. \ref{complexityanalysis}).
\vspace{-0.01cm}
\begin{table}[!ht]
\begin{center}
\caption{The parameters used in Algorithm \ref{alg:B} with level $=4$}
\label{table2}
\begin{tabular}[h]{|c|c|c|c|c|c|}
\hline
\multicolumn{6}{|c|}{$N=1024,K=512$}\\
\hline
$E_b/N_0$(dB) & 1.5 & 1.75 &2 &2.25&2.5 \\
\hline
$(\gamma_{\text{left}},\gamma_{\text{right}})$ &(3.6, 2)&(3.6, 2)&(3.6, 2)&(4, 3)&(6, 5)\\
\hline
$\omega_0, \omega_1, \omega_4$ & $\varnothing$ & $\varnothing$ & $\varnothing$ & $\varnothing$ & $\varnothing$ \\
\hline
$\omega_2$ & 0.5& 0.5& 0.5& 0.6&0.6\\
\hline
$\omega_3$ & 0.25& 0.25 & 0.25 &  0.3 &0.3\\
\hline
\end{tabular}
\end{center}
\vspace{-0.65cm}
\end{table}

The average computational complexity of Algorithm \ref{alg:B} is investigated in Fig. \ref{complexityanalysis}. It can be seen that the average complexity decreases rapidly as SNR increases. The reason is that as the underlying channel turns to be more reliable, it is sufficient to flip only one or two unfrozen bits to obtain the correct estimate for most cases, and the search stops at an early time. Note that, in low SNR regime, the complexity of the proposed algorithm grows rapidly since more paths need to be searched when error becomes more random, while we also note that in practical system, the low SNR region is not a region of interest because the decoding procedure is usually not activated at a low SNR region due to the high BLER.

\section{conclusion}
A critical set $S$ which with high probability includes the first incorrect hard decision in SC decoding is proposed. By iteratively modifying the critical set, multi-layer critical sets are established. On this basis, a progressive multi-level bit-flipping decoder which can correct multiple errors in SC decoding is proposed.
We show that as the level increases, the BLER performance of the proposed progressive bit-flipping decoder competes with the corresponding CA-SCL decoder. Furthermore, the average complexity of the proposed algorithm is much lower than that of a SCL decoder (and is similar to that of SC decoding) at medium to high SNR.

\section*{Acknowledgement}
This work was supported in part by National Hi-Tech R\&D Program of China (No. 2014AA01A702), National Natural Science Foundation of China (No. 61371094, No. 61401391), National Key Basic Research Program of China (No. 2012CB316104), Zhejiang Provincial Natural Science Foundation (No. LR12F01002), the open project of Zhejiang Provincial Key Laboratory of Information Proc., Commun. \& Netw., China, HIRP Flagship Projects from Huawei Technologies Co., Ltd (YB2013120029 and YB2015040053), Natural Science Foundation of Fujian Province (No. 2017J01106), and Key Project of Natural Science Fund for Young Scholars in Universities and Colleges of Fujian Province (No. JZ160489).

\bibliographystyle{IEEEtran}

\end{document}